\documentclass[runningheads,a4paper]{llncs}
\usepackage{amsmath}
\usepackage{amssymb}
\usepackage{mathtools}

\usepackage{algorithm,algorithmic}

\usepackage{pgfplots}
\usepackage{subfig}
\usepackage{subfig}
\usepackage{setspace}
\usepackage{cite}
\usepackage{graphicx}
\usepackage{color}
\usepackage[strings]{underscore}

\usepackage{tikz}
\usetikzlibrary{arrows,decorations.markings,plotmarks}
\usetikzlibrary{calc}
\usetikzlibrary{positioning}
\usetikzlibrary{backgrounds,positioning}
\usetikzlibrary{decorations.pathreplacing}

\usepackage{mathtools}
\usepackage{color}
\usepackage{amsmath}

\usepackage{graphicx}
\usepackage{color}
\usepackage[strings]{underscore}

\newcommand{\blu}{\textcolor{blue}}
\newcommand{\magenta}{\textcolor{magenta}}
\newcommand{\green}{\textcolor{green}}

\begin{document}

\title{Bayesian Verification\\ of Chemical Reaction Networks}
%
%
\author{Gareth W. Molyneux \thanks{Gareth Molyneux acknowledges funding from the University of Oxford, the EPSRC \& BBSRC Centre for Doctoral Training in Synthetic Biology (grant EP/L016494/1).} \and
Viraj B. Wijesuriya \and
Alessandro Abate}
\authorrunning{G. W. Molyneux et al.}
\institute{Department of Computer Science, University of Oxford, Oxford, OX1 3QD, UK
\email{{firstname}.{lastname}@cs.ox.ac.uk}}
\maketitle              

\begin{abstract}
We present a data-driven verification approach that determines whether or not a given chemical reaction network (CRN) satisfies a given property,
expressed as a formula in a modal logic.
Our approach consists of three phases, integrating formal verification over models with learning from data.
First, we consider a parametric set of possible models based on a known stoichiometry and classify them against the property of interest.
Secondly, we utilise Bayesian inference to update a probability distribution of the parameters within a parametric model with data gathered from the underlying CRN.
In the third and final stage, we combine the results of both steps to compute the probability that the underlying CRN satisfies the given property.
We apply the new approach to a case study and compare it to Bayesian statistical model checking. 
\end{abstract}

\section{Introduction}

Constructing complete models of biological systems with a high degree of accuracy is a prevalent problem in systems and synthetic biology.
Attaining full knowledge of many existing biological systems is impossible, making their analysis, prediction, and the designing of novel biological devices an encumbrance.
In this work, we integrate the use of probabilistic model-based analysis techniques with a data-based approach via Bayesian inference.
Chemical Reaction Networks (CRNs) \cite{Cook2009} provide a convenient formalism for describing various biological processes as a system of well-mixed reactive species in a volume of fixed size.
This methodology allows for the construction of an accurate model from the data to verify that the underlying data-generating system satisfies a given formal property.
Thus, by verifying the properties of the model, we can assert quantatively whether the underlying data generating system satisfies a given property of interest.
We leverage model analysis by means of formal verification, namely quantitative model checking~\cite{Baier:2008:PMC:1373322}.
The end result is the computation of a probability, based on the collected data, that the underlying system satisfies a given formal specification.
If the obtained probability is closer to either one or zero, we can confidently draw an assertion on the satisfaction of the property over the underlying biological system.
On the other hand, with a moderate probability value, a decision on the experimental setup or on the models needs to be made: we can either collect more data from the experiments, or propose alternative models and start the procedure once more.
The proposed approach is different from statistical model checking (SMC) \cite{Agha2018}, in that standard SMC procedures require target systems with fully known models: these are also in general too large for conventional probabilistic model checkers (PMC) \cite{Baier:2008:PMC:1373322}.
Alternative SMC procedures can also work with unknown models, provided that one is able to produce fully observable traces.
Our work instead targets partially known systems that produce noisy observations at discrete points in time, which are commonplace in biology: these systems are captured by a parametric model class with imperfect knowledge of rates within a known stoichiometry.
The new approach comprises of three phases.
First, we propose a parametric model of a given, partially known biological system, and perform parameter synthesis\cite{Ceska2014} to determine a set of parameters over the parametric models that relates to models verifying the given property.
This is performed via PRISM \cite{Ceska2014,Ceska2016}.
The second phase, executed in parallel with the first, uses Bayesian inference to infer posterior distributions over the likely values of the parameters, based on data collected from the underlying partially known and discretely observed system.
In the third phase, we combine the outputs from the two phases to compute the probability that the model satisfies the desired property, which results in an assertion on the satisfaction of the property over the underlying biological system.

\noindent\emph{\textbf{Related Work}}
CRNs have been utilised to model biological systems both deterministically \cite{Angeli2009} and stochastically \cite{Wilkinson2011} via the chemical master equation \cite{Gillespie1992}.
We use continuous-time Markov chains\cite{Karlin1975} (CTMCs) to model CRNs.
Both probabilistic model checking approaches\cite{Baier2003,Kwiatkowska2007} and statistical model checking approaches \cite{Agha2018} have been applied in many areas within biology \cite{Kwiatkowska2014,Kwiatkowska2017,Zuliani2015} with tools such as PRISM \cite{Kwiatkowska2011}, providing crucial support to perform procedures for continuous-time Markov chains such as parameter synthesis\cite{Ceska2014,Brim2013,Han2008}.
Bayesian inference\cite{Box1973,Broemeling2017} techniques have long been applied to biological systems\cite{Lawrence2010}.
In particular, we focus on inferring the kinetic parameters of the CRNs\cite{Schnoerr2017,Wilkinson2010,Boys2008}.
Exact inference is difficult due to the intractability of the likelihood function.
Sampling techniques such as particle Markov chain Monte Carlo \cite{Golightly2006,Golighty2011} and likelihood-free methods\cite{Murphy2012,Revell2018} such as approximate Bayesian computation\cite{Toni2008,Warne2019} have been utilised to circumvent intractable likelihoods.
Inferring parameters and formally verifying properties using statistical model checking for deterministic models is considered in \cite{Gyori2014}.
Computing probability estimates using data produced by an underlying stochastic system, driven by external inputs to satisfy a given property, is considered in \cite{Haesaert2015}.
The integration of the parameter synthesis problem and Bayesian inference is considered for discrete-time Markov chains in \cite{Polgreen2016a} with the extension to actions for Markov decision processes in \cite{Polgreen2017a}.
In \cite{Polgreen2016a}, the authors consider exact parameter inference for a discrete state, discrete time system that consists of a handful of states with fully observed, continuous data.
In our work, the data considered are discretely observed data points produced by a single simulation from a continuous-time Markov chain given the true parameters, which is then perturbed by noise and we pursue likelihood free inference in the form of approximate Bayesian Computation\cite{Beaumont2002,Sisson2018}.
Our approach is then compared to a Bayesian approach to statistical model checking\cite{Zuliani2010,Zuliani2009}.

The problem of learning and designing continuous-time Markov chains subject to the satisfaction of properties is considered in \cite{Bortolussi2013} meanwhile the  model checking problem is reformulated to a sequential Bayesian computation of the likelihood of an auxiliary observation process in \cite{Milios2017}.
Directly related work is presented in \cite{Bortolussi2016}; a Bayesian statistical algorithm was developed that defines a Gaussian Process (GP) \cite{Rasmussen2003} over the parameter space based on a few observations of true evaluations of the satisfaction function. 
The authors build upon the idea presented in \cite{Bortolussi2013} and define the satisfaction function as a smooth function of the uncertain parameters of a given CTMC, where this smooth function can be approximated  by a GP.
This GP allows one to predict the value of the satisfaction probability at every value of the uncertain parameters from individual model simulations at a finite number of distinct parameter values.
This model checking approach is incorporated into the parameter synthesis problem considered in \cite{Bortolussi2018} which builds upon the parameter synthesis problem defined in \cite{Ceska2016}, but differs with the incorporation of the model checking approach presented in \cite{Bortolussi2016} and an active learning step being introduced to adaptively refine the synthesis.
Model construction and selection via Bayesian design is presented in \cite{Barnes2011a,Barnes2011b,Woods2016}.

The rest of the paper is as follows.
In Section 2, we cover the necessary background material required for our framework.
In Section 3, we introduce our framework, covering parameter synthesis, Bayesian inference and the probability calculation techniques required.
In Section 4, we consider the application of this framework to a case study and compare our framework to Bayesian statistical model checking\cite{Zuliani2010}.
We conclude with a discussion of our work and possible extensions.
\section{Background}

\subsection{Parametric Continuous-Time Markov Chains} 
We work with discrete-state, continuous-time Markov chains\cite{Karlin1975}.

\begin{definition}
[Continuous-time Markov Chain]

\label{CTMC}
A continuous-time Markov chain (CTMC) $\mathcal{M}$ is a tuple $(\mathcal{S},\textbf{R},AP,L)$, where;
\begin{itemize}
\item{$\mathcal{S}$ is a finite, non-empty set of states,}
\item{$s_0$ is the initial state of the CTMC,}
\item{$\textbf{R} : \mathcal{S} \times \mathcal{S} \rightarrow \mathbb{R}_{\geq 0}$ is the transition rate matrix, where $\textbf{R}(s,s')$ is the rate of transitioning from state $s$ to state $s'$,}
\item{$L : \mathcal{S} \rightarrow 2^{AP}$ is a labelling function mapping each state, $s \in S$, to the set $L(s) \subseteq AP$ of atomic propositions $AP$, that hold true in $s$.}
\end{itemize}
\end{definition}
The transition rate matrix $\textbf{R}$ governs the dynamics of the overall model.
A transition between states $s$ and $s'$ can only occur if $\textbf{R}(s,s')>0$ and $s \neq s'$, in which case, the probability of triggering the transition within a time $t$ is $1 - e^{-t \textbf{R}(s,s')}$.
If $s = s'$, $\textbf{R}(s,s) = -E(s) = -\sum_{s' \in \mathcal{S}} \textbf{R}(s,s')$, where $E(s)$ is defined as the exit rate from $s$.
The time spent in state $s$ before a transition is triggered is exponentially distributed by the exit rate, $E(s)$. 
%
We define a sample trajectory or path of a CTMC as follows. 
\begin{definition}
[Path of a CTMC]

Let $\mathcal{M} = (\mathcal{S},\textbf{R},AP,L)$ be a CTMC.
A path $\omega$ of $\mathcal{M}$ is a sequence of states and times $\omega = s_0 t_0 s_1 t_1 \dots,$ where for all $i = 0,1,2,\dots,n,$, $s_i \in \mathcal{S}$ and $t_i \in \mathbb{R}_{\geq 0}$, is the time spent in state $s_i$.
\end{definition}

\smallskip

Parametric continuous-time Markov chains (pCTMCs) extend the notion of CTMCs by allowing transition rates to depend on a vector of model parameters, $\boldsymbol{\theta} = (\theta_1, \theta_2,...,\theta_k)$.
The domain of each parameter $\theta_k$ is given by a closed real interval describing the range of possible values, $[\theta_k^{\bot},\theta_k^{\top}]$.
The parameter space $\Theta$ is defined as the Cartesian product of the individual intervals, $\Theta = \bigtimes_{\tilde{k}\in\lbrace 1,\dots,k \rbrace}[\theta_{\tilde{k}}^{\bot},\theta_{\tilde{k}}^{\top}]$, so that $\Theta$ is a hyperrectangular set. 
\begin{definition}
[Parametric CTMC]

Let $\Theta$ be a set of model parameters.
A parametric Continuous-time Markov Chain (pCTMC) over $\boldsymbol{\theta}$ is a tuple $(\mathcal{S},\textbf{R}_{\boldsymbol{\theta}},AP,L)$, where:
\begin{itemize}
\item{$\mathcal{S},s_0,AP$ and $L$ are as in Definition \ref{CTMC}, and}
\item{$\boldsymbol{\theta} = (\theta_1, \dots, \theta_k)$ is the vector of parameters, taking values in a compact hyperrectangle $\Theta \subset \mathbb{R}^k_{\geq 0}$,}
\item{$\textbf{R}_{\boldsymbol{\theta}}:\mathcal{S}\times \mathcal{S}\rightarrow \mathbb{R}[\boldsymbol{\theta}]$ is the parametric rate matrix, where $\mathbb{R}[\boldsymbol{\theta}]$ denotes a set of polynomials over the reals $\mathbb{R}$ with variables $\theta_k$, $ \boldsymbol{\theta} \in \Theta$.}
\end{itemize}
\end{definition}

Given a pCTMC and a parameter space $\Theta$, we denote with $\mathcal{M}_{\boldsymbol{\Theta}}$ the set $\lbrace \mathcal{M}_{\boldsymbol{\theta}} | \boldsymbol{\theta} \in \Theta \rbrace$ where $\mathcal{M}_{\boldsymbol{\theta}} = (\mathcal{S},\textbf{R}_{\boldsymbol{\theta}},AP,L)$ is the instantiated CTMC obtained by replacing the parameters in $\textbf{R}$ with their valuation in $\boldsymbol{\theta}$.
We restrict the rates to be polynomials, which are sufficient to describe a wide class of biological systems\cite{Gillespie1977b}.

\subsection{Properties - Continuous Stochastic Logic}

We aim to verify properties over pCTMCs.
To achieve this, we employ the time-bounded fragment of \emph{continuous stochastic logic} (CSL) \cite{Aziz1996,Kwiatkowska2007}.

\begin{definition}
Let $\phi$ be a CSL formula interpreted over states $s\in \mathcal{S}$ of a pCTMC $\mathcal{M}_{\boldsymbol{\theta}}$,
and $\varphi$ be a formula over its paths.
The syntax of CSL is given by
\begin{equation*}
\phi := \text{true} \ |\ a \ |\ \neg \phi\ |\ \phi \wedge \phi \ |\ \phi \vee \phi \ |\ P_{\sim p}[ \varphi ]
\end{equation*}
\begin{equation*}
\varphi := X\phi \ | \ \phi U^{[t,t']} \phi \ | \ \phi U \phi,
\end{equation*}
where $a \in AP$, $\sim \ \in \lbrace <, \leq, \geq, > \rbrace$, $p \in [0,1]$, and $t,t' \in \mathbb{R}_{\geq 0}$. \\
\end{definition}
$P_{\sim p}[\varphi]$ holds if the probability of the path formula $\varphi$ being satisfied from a given state meets $\sim p$.
Path formulas are defined by combining state formulas through temporal operators: $X\phi$ is true if $\phi$ holds in the next state, $\phi_1 U^{I} \phi_2$ is true if $\phi_2$ holds at all time points $t \in I$ and $\phi_1$ holds for all time points $t' < t$.
%
We now define a \emph{satisfaction function} to capture how the satisfaction probability of a given property relates to the parameters and the initial state.
\begin{definition}
[Satisfaction Function]
\label{SatFunc}
Let $\phi$ be a CSL formula, $\mathcal{M}_{\boldsymbol{\theta}}$ be a pCTMC over a space $\Theta$,  $s \in \mathcal{S}$, $s_0$ is the initial state, and $Path^{\mathcal{M}_{\boldsymbol{\theta}}}(s_0)$ is the set of all paths generated by $\mathcal{M}_{\boldsymbol{\theta}}$ with initial state $s_0$.
Denote by $\Lambda_{\phi}:\boldsymbol{\theta} \rightarrow [0,1]$ the satisfaction function such that $\Lambda_{\phi}(\boldsymbol{\theta})= Prob\left(\lbrace \omega \in Path^{\mathcal{M}_{\boldsymbol{\theta}}}(s_0) \models \varphi \rbrace \ | \omega(0) = s_0 \right) \sim p$.
\end{definition}
That is, $\Lambda_{\phi}(\boldsymbol{\theta})$ is the probability that a pCMTC $\mathcal{M}_{\boldsymbol{\theta}}$ satisfies a property $\phi$, $\mathcal{M}_{\boldsymbol{\theta}} \models \phi $.

\subsection{Stochastic modelling of Chemical Reaction Networks}
\label{CRN-CTMC}
Semantics for continuous-time Markov chains include states that describe the number of molecules of each species and transitions which correspond to reactions that consume and produce molecules.
These reactions are typically parameterised by a set of kinetic parameters that dictate the dynamics of the overall network and it is these parametric CRNs that we will turn our focus towards:
\begin{definition}
[Parametric Chemical Reaction Network ]
A parametric Chemical Reaction Network (pCRN) $\mathcal{C}$ is a tuple $(M,\textbf{X},W,\mathcal{R},\boldsymbol{\upsilon})$ where
\begin{itemize}
\item{$M = \lbrace m_1,\ldots,m_n\rbrace$ is the set of n species; }
\item{$\textbf{X} = (X_1,...,X_n)$ is a vector where each $X_i$ represents the number of molecules of each species $i \in \lbrace 1,...,n\rbrace$.  $\textbf{X} \in W$, with $W \subseteq \mathbb{N}^{N}$ the state space;}
\item{$\mathcal{R} = \lbrace r_1, \ldots, r_k\rbrace$ is the set of chemical reactions, each of the form $r_j = (\textbf{\text{v}}_j, \alpha_j)$, with $\textbf{v}_j$ the stoichiometry vector of size $n$ and $\alpha_j = \alpha_j(\textbf{X},\upsilon_j)$ is the propensity or rate function.}
\item{$\boldsymbol{\upsilon} = (\upsilon_1,\ldots,\upsilon_k)$ is the vector of (kinetic) parameters, taking values in a compact hyper-rectangle $\Upsilon \subset \mathbb{R}^k$.}
\end{itemize}
\end{definition}

Each reaction $j$ of the pCRN can be represented as
\begin{equation}
r_j : u_{j,1} m_1 + \ldots + u_{j,n} m_n \xrightarrow{\alpha_j} u'_{j,1}m_1 + \ldots + u'_{j,n} m_n,
\end{equation}
where $u_{j,i}$ $(u'_{j,i})$ is the amount of species $m_i$ consumed (produced) by reaction $r_j$.
The stoichiometric vector $\textbf{v}_j$ is defined by $\textbf{v}_j = \textbf{u$^{\prime}$}_j - \textbf{u}_j$, where $\textbf{u}_j = (u_{j,1},\dots, u_{j,n})$ and $\textbf{u$'$}_j = (u'_{j,1},\dots, u'_{j,n})$.

A pCRN can be modelled as a pCTMC if we consider each state of the pCTMC to be a unique combination of the number of molecules.
That is, if we denote $\textbf{X}(t_i)$ as the number of molecules of each species at a given time, $t_i$, then the corresponding state of the pCTMC at time $t_i$ is $s_i = \textbf{X}(t_i)$.
In fact, pCTMC semantics can be derived such that the transitions in the pCTMC correspond to reactions that consume and produce molecules, by defining the rate matrix as:
\begin{equation}
\textbf{R}(s_i,s_j) = \sum_{j \in \zeta(s_i,s_j)}\alpha_j(s_i,\upsilon_j) = \sum_{j \in \zeta(s_i,s_j)}\upsilon_j g_j(s_i),
\end{equation}
where $\zeta(s_i,s_j)$ denotes all the reactions changing state $s_i$ into $s_j$ and $\alpha_j$ is the propensity or rate function defined earlier and the propensity, $\alpha_j$, often takes the form $\alpha_j(s_i,\upsilon_j) = \upsilon_j g_j(s_i)$, where $g_j(s_i)$ is the combinatorial factor that is determined by the number of molecules in the current state, $s_i$ and the type of reaction $j$.
It is clear to see that this new pCTMC is governed by the kinetic rate parameters, $\boldsymbol{\upsilon}$, thus, $\mathcal{M}_{\boldsymbol{\upsilon}}$ is the pCTMC that models the pCRN and for the rest of this paper, with a slight abuse in notation, we will let $\mathcal{M}_{\boldsymbol{\theta}}$ be the pCTMC that represents a pCRN where $\boldsymbol{\theta}$ are the kinetic rate parameters.
Now the vector of kinetic parameters is defined as $\boldsymbol{\theta} = (\theta_1,\ldots,\theta_k)$, where $\boldsymbol{\theta} \in \Theta$ and $\Theta \subset \mathbb{R}^k$.

\subsection{Bayesian Inference}

When constructing mathematical models to describe real applications, statistical inference is performed to estimate the model parameters from the observed data.
Bayesian inference \cite{Box1973} is performed by working either with or without a parametric model and experimental data, utilising the experimental data available to approximate the parameters in a given model and to quantify any uncertainties associated with the approximations.
It is of particular interest to the biological community to constrain any uncertainty within the model parameters (or indeed the model itself) by using the observed data of biological systems.
Moreover, when one is working with obstreperous stochastic models, noisy observations may add another layer of uncertainty.
A plethora of literature is focused on the problem of Bayesian inference in stochastic biochemical models\cite{Wilkinson2011,Schnoerr2017,Warne2019}, let alone stochastic models \cite{Broemeling2017}.
Bayesian methods have been used extensively in the life sciences for parameter estimation, model selection and even the design of experiments \cite{Toni2008,Liepe2013,Vanlier2014,Galagali2014,Sanguinetti2006,Lawrence2010}.

Given a set of observations or data, $D$, and a model governed by $\boldsymbol{\theta}$, the task of Bayesian inference is to learn the true parameter values given the data and some existing prior knowledge.
This is expressed through Bayes' theorem:
\begin{equation}
\label{Bayes}
p(\boldsymbol{\theta} | D) = \frac{p(D\ | \ \boldsymbol{\theta})p(\boldsymbol{\theta})}{p(D)}.
\end{equation}
Here $p(\boldsymbol{\theta}|D)$ represents the \emph{posterior} distribution, which is the probability density function for the parameter vector, $\boldsymbol{\theta}$, given the data, $D$; $p(\boldsymbol{\theta})$ is the \emph{prior} probability distribution which is the probability density of the parameter vector before considering the data; $p(D | \boldsymbol{\theta})$ is the \emph{likelihood} of observing the data given a parameter combination; and $p(D)$ is the \emph{evidence}, that is, the probability of observing the data over all possible parameter valuations.
Assumptions about the parameters are encoded in the prior meanwhile assumptions about the model itself are encoded into the likelihood.
The evidence acts as a normalisation constant and ensures the posterior distribution is a proper probability distribution.
To estimate the posterior probability distribution, we will utilize Monte Carlo techniques.

\section{Bayesian Verification}

The main problem we address in this work is as follows.
Consider a real-life, data generating biological system $\textbf{S}$, where we denote the data generated by the system as $D$ and we are interested in verifying a property of interest, say $\phi$.
Can we use this obtained data and the existing knowledge of the model to formally verify a given property over this system, $\textbf{S}$?

\begin{figure}
 \begin{center}

\begin{tikzpicture}[every node/.style={minimum height={1.0cm},thick,align=center}]

\node[draw] (Prop) {Property, \magenta{$\phi$}};

\node[draw,  right = 0.35cm of Prop] (Model) {Model Class
\\
(pCTMC) $\mathcal{M}_{\blu{\boldsymbol{\theta}}}$};

\node[draw, right = 2cm of Model] (Data) {Generate data, \green{$D$} from
\\
System, \textbf{S} };

\node[draw,  below left  = 0.5cm and -2cm of Model] (Synth) {Parameter
Synthesis
\\
$\blu{\Theta}_{\magenta{\phi}} = \lbrace \blu{\boldsymbol{\theta}} \in \blu{\Theta} : \mathcal{M}_{\blu{\boldsymbol{\theta}}} \models \magenta{\phi}\rbrace \subseteq \blu{\Theta}$};

\node[draw, right = 2.5cm of Synth] (Inf) {Bayesian Inference,
\\
$p(\blu{\boldsymbol{\theta}} | \green{D})$};

\coordinate (CENTER) at ($(Inf)!0.5!(Synth)$);
\node[draw,below = 1.0cm of CENTER] (Confidence) {Probability
\\
Calculation
\\
$\mathcal{C} = P(\textbf{S} \models \magenta{\phi} | \green{D})  = \int_{\blu{\Theta}_{\magenta{\phi}}} p(\blu{\boldsymbol{\theta}} | \green{D}) d\blu{\boldsymbol{\theta}}$};

\draw[->] (Prop.south) to [out = 270,in = 90] (Synth.north);
\draw[->] (Model.south) to [out = 235, in = 90] (Synth.north);

\draw[->] (Synth.east) to [out = 0, in = 90]  (Confidence.90);
\draw[->] (Data.south) -- (Inf.north);
\draw[->] (Inf.west)  to [out = 180, in = 90] (Confidence.90);
\end{tikzpicture}
\end{center}
\caption{Bayesian Verification Framework. \label{BayVer}}
 \end{figure}
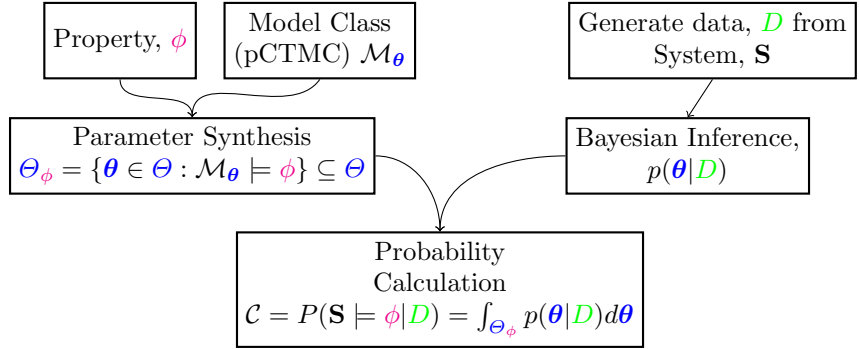

Here on, we will be considering this problem using chemical reaction networks to describe biological systems.
We assume that we have sufficient knowledge to propose a parametric model for the underlying system, which in this case is a pCTMC denoted by $\mathcal{M}_{\boldsymbol{\theta}}$.
We define the property of interest, $\phi$, in CSL and we also assume that we are able to obtain data, $D$, from the underlying system.
There are three aspects to the Bayesian Verification framework: parameter synthesis\cite{Ceska2014,Ceska2016}, Bayesian inference \cite{Box1973,Murphy2012,Broemeling2017} and a probability or credibility interval calculation \cite{Box1973}.
We shall discuss the data we work with and these methods in detail later.
Given a model class $\mathcal{M}_{\boldsymbol{\theta}}$ and a property of interest, $\phi$, we first synthesise a set of parameter valuations $\Theta_{\phi} \subseteq \Theta$.
If we were to choose a vector of parameters $\boldsymbol{\theta}'$ such that $\boldsymbol{\theta}' \in \Theta_{\phi}$, then the paths or traces generated from the induced pCTMC, $\mathcal{M}_{\boldsymbol{\theta'}}$ would satisfy the property of interest with some probability, which we denote as $\mathcal{M}_{\boldsymbol{\theta}'} \models \phi$.
We learn the parameters of interest by inferring them from the data via Bayesian inference, to provide us with a posterior distribution, $p(\boldsymbol{\theta} | D ) $.
Once we have this posterior distribution and a synthesised set of parameter regions, $\Theta_{\phi}$ we integrate the posterior probability distribution over these regions to obtain a probability on whether the underlying data generating system satisfies the property or not.
The full procedure is illustrated in Figure \ref{BayVer}.

\subsection{Parameter Synthesis}
\label{ParamSynthesis}
Given a parametric model class $\mathcal{M}_{\boldsymbol{\theta}}$ and a property $\phi$ defined in CSL, we synthesise parameter regions that satisfy $\phi$ using the approach introduced in \cite{Ceska2016}.
We will focus on the threshold synthesis problem.
Note that solutions to the threshold synthesis problem may sometimes lead to parameter points that are left undecided, that is, parameter points that either do or do not satisfy the property with a given probability bound, $\sim p$.
Let us define this problem formally.
\begin{definition}
[Threshold Synthesis]
Let $\mathcal{M}_{\boldsymbol{\theta}}$ be a pCTMC over a parameter space $\Theta$, $\phi$ a CSL formula, $\sim p$ a threshold where $p \in [0,1]$, $\sim \in \lbrace \leq, <,>,\geq \rbrace$ and $\mathcal{E} > 0 $ be a volume tolerance.
The threshold synthesis problem is finding a partition $\lbrace \mathcal{T}, \mathcal{U}, \mathcal{F}\rbrace$ of $\Theta$ such that:
\begin{enumerate}
\item{$\forall \boldsymbol{\theta} \in \mathcal{T}$. $\Lambda_{\phi}(\boldsymbol{\theta}) \sim p$; and}

\item{$\forall \boldsymbol{\theta} \in \mathcal{F}$. $\Lambda_{\phi}(\boldsymbol{\theta}) \nsim p$; and}

\item{$vol(\mathcal{U})/vol(\Theta) \leq \mathcal{E}$}
\end{enumerate}
where $vol(A) $ is the volume of $A$.
\end{definition}

%

The goal of parameter synthesis is to synthesise the set of all possible valuations for which the model class $\mathcal{M}_{\boldsymbol{\theta}}$ satisfies the property $\phi$:
\begin{equation}
\Theta_{\phi} = \lbrace \boldsymbol{\theta} \in \Theta \ : \ \mathcal{M}_{\boldsymbol{\theta}} \models \phi \rbrace.
\end{equation}
We define the region $\Theta_{\phi} \subseteq \Theta$ as the feasible set of parameters.
Parametric model checking capabilities of the tool introduced in \cite{Ceska2016} is leveraged to perform parameter synthesis over the CTMC constructed from a given pCRN.

\subsection{Bayesian Inference for Parametric CTMC}\label{BayesianInf}

In this section, we discuss the application of Bayesian inference for parametric CTMCs to infer unknown model parameters.
Inferring parameters from pCTMCs is a widely studied problem in the realms of biology \cite{Wilkinson2011,Schnoerr2017,Golighty2011,Warne2019,Toni2008,
Wilkinson2010,Georgoulas2017,Golighty2015,Boys2008}.
The focus of our work here will be on performing inference over noisy time series data  that has been observed a finite number of times at discrete points in time.

\subsubsection{Partially observed data }
\label{PartiallyObservedData}
Let us consider the case where the data $D$ consists of $Q$ observations of the CRN state vector at discrete points in time, $\tilde{t}_1,\tilde{t}_2,\dots, \tilde{t}_Q$.
Let $D = [\textbf{Y}(\tilde{t}_1),\textbf{Y}(\tilde{t}_2),\dots, \textbf{Y}(\tilde{t}_Q)]$, where $\textbf{Y}(\tilde{t}_i)\in \mathbb{R}^N$ represents an observation of the molecule count sample $\textbf{X}(\tilde{t}_i)$, which has a corresponding state $s_i$ in the pCTMC.
It is common to incorporate uncertainty in these observations with the use of additive noise\cite{Schnoerr2017,Wilkinson2011},
\begin{equation}
\label{Obs}
\textbf{Y}(\tilde{t}_i) = \textbf{O}\textbf{X}(\tilde{t}_i) + \boldsymbol{\xi},
\end{equation}
where $\textbf{O}$ is a $O \times n$ matrix, $\mathbb{R}^{O \times n}$ and $\boldsymbol{\xi}$ is a $O \times 1$ vector of independent Gaussian random variables.
The observation vectors $\textbf{Y}(\tilde{t}_i)$, are  $O \times 1$ vectors where  $O \leq n$, which reflects the fact that only a sub-set of chemical species of $\textbf{X}(t_i)$ are observed.
For this work, $\textbf{O} = \textbf{I}$, where $\textbf{I}$ is an $n \times n$ identity matrix, recalling that $n$ is the number of different chemical species.
Due to both the nature of data we are working with and the intractability of the chemical master equation\cite{Gillespie1992} that determines the likelihood, we turn away from working with the analytical likelihood to consider likelihood free methods\cite{Murphy2012,Warne2019}.
Two popular classes of likelihood-free inference methods available are pseudo-marginal Markov chain Monte Carlo\cite{Andrieu2009} and Approximate Bayesian Computation (ABC)\cite{Beaumont2002,Sisson2018}.
In our work, we utilise ABC to infer parameters of our model.
Not only do ABC methods allow working with highly complicated models with intractable likelihoods to be investigated, but also ABC methods are very intuitive and easy to implement - it has proven to be an invaluable tool in the life sciences \cite{Beaumont2002,Kypraios2017,Toni2008,Beaumont2010}.
To deploy ABC methods, we need to be able to simulate trajectories from a given model of interest, which in our case is a pCTMC, and require a discrepancy metric, $\rho ( D, \tilde{\textbf{X}})$, where $\tilde{\textbf{X}} = (\tilde{\textbf{X}}(\tilde{t}_1),\dots,\tilde{\textbf{X}}(\tilde{t}_M))$ is the vector of simulated data generated through the model that consists of $M$ reactions.
This discrepancy metric provides a measure of distance between that of the experimental data and the simulated data and this simulated data will form the basis of our Bayesian inference technique.
After calculating $\rho(D,\tilde{\textbf{X}})$, we accept the traces where $\rho(D,\tilde{\textbf{X}}) \leq \boldsymbol{\epsilon}$, where $\boldsymbol{\epsilon}$ is the discrepancy threshold.
This leads to a modification of the original Bayes theorem 
\begin{equation}
\label{ABC} p(\boldsymbol{\theta} | \rho(D,\tilde{\textbf{X}}) \leq \boldsymbol{\epsilon}) =  \frac{p( \rho ( D, \tilde{\textbf{X}}) \leq \boldsymbol{\epsilon} \ | \ \boldsymbol{\theta}) p(\boldsymbol{\theta})}{p(\rho(D, \tilde{\textbf{X}}) \leq \boldsymbol{\epsilon}) }.
\end{equation}
For the prior probability distribution, $p(\boldsymbol{\theta})$, we will assume a uniform prior over the possible parameter set, $\boldsymbol{\theta}$.
By being able to produce simulations from the model, we are able to perform inference for the parameters of interest, subject to data $D$.
The discrepancy threshold $\boldsymbol{\epsilon}$ determines the level of approximation - as $\boldsymbol{\epsilon} \rightarrow 0$, $p(\boldsymbol{\theta} \ | \ \rho ( D, \tilde{\textbf{X}}) \leq \boldsymbol{\epsilon}) \rightarrow p(\boldsymbol{\theta} | D)$.
In practice, Equation \eqref{ABC} can be treated as an exact posterior under the assumption of model and observation error when $\boldsymbol{\epsilon} \rightarrow 0$\cite{Wilkinson2013}.
Picking an appropriate discrepancy metric is a challenge in itself\cite{Sisson2018} as the choice in discrepancy metric can lead to bias.
The discrepancy metric used in our work is defined by
\begin{equation}
\label{DiscrepancyMet}
\rho(D, \tilde{\textbf{X}}) = \left[ \sum^Q_{i=1} (\textbf{Y}(\tilde{t}_i) - \tilde{\textbf{X}}(\tilde{t}_i))^2 \right]^{1/2},
\end{equation}
Clearly for any $\boldsymbol{\epsilon} > 0$, ABC methods produce biased results and this bias should be considered in any subsequent results we obtain, especially for any Monte Carlo estimate.
In order to estimate integrals such as the expected mean and covariance, which is necessary for the posterior probability distribution, we must be able to generate samples, $\boldsymbol{\theta}^{(i)}$ from the posterior.
A summary of different methods available to obtain these samples can be found in \cite{Warne2019} along with a detailed discussion on every method.
We will be focusing on the approximate Bayesian computation sequential Monte Carlo (ABCSeq) approach \cite{Sisson2007,Beaumont2009,Toni2008}.
The idea behind the ABCSeq approach is to use sequential importance resampling to propogate $m$ samples, called particles, through a sequence of $R + 1$ ABC posterior distributions defined through a sequence of discrepancy thresholds, $\boldsymbol{\epsilon}_0,\boldsymbol{\epsilon}_1,\dots,\boldsymbol{\epsilon}_R$, with $\boldsymbol{\epsilon}_r > \boldsymbol{\epsilon}_{r+1}$, for $r = 0,1,\dots, R-1$, for a number of $R$ thresholds and $\boldsymbol{\epsilon}_0 = \infty$.
The method is presented in Algorithm \ref{ABCSMCAlg}.
\begin{minipage}{\linewidth}

\begin{algorithm}[H] 
\scriptsize
\begin{algorithmic}[1] 
    \STATE Initialize threshold sequence $\boldsymbol{\epsilon}_0 > \cdots > \boldsymbol{\epsilon}_R$
    \STATE Set {$r = 0$}
    	\FOR{$i = 1,\dots,m$}
    	\STATE Simulate $\boldsymbol{\theta}^{(0)}_{i} \sim p(\boldsymbol{\theta})$ and $\tilde{\textbf{X}} \sim p(\tilde{\textbf{X}} | \theta_i^{(0)})$ until $\rho(D,\tilde{\textbf{X}}) < \boldsymbol{\epsilon}_1$
    	\STATE $w_i = 1 / m$
    	\ENDFOR
    \FOR{$r = 1,\dots,R-1$}
    	\FOR{$i = 1,\dots, m$}
    		\WHILE{ $\rho(D,\tilde{\textbf{X}}) > \boldsymbol{\epsilon}_r$}
    			\STATE Pick $\boldsymbol{\theta}^*_{i}$ from the previously sampled $\boldsymbol{\theta}_i^{(r-1)}$ with corresponding probabilities $w_i^{(r-1)}$, draw $\boldsymbol{\theta}_i^{(r)} \sim K_r(\boldsymbol{\theta}_i^{(r)} | \boldsymbol{\theta}_i^*)$ and $\tilde{\textbf{X}} \sim p( \tilde{\textbf{X}} | \boldsymbol{\theta}_i^{(r)})$
    		\ENDWHILE
    		\STATE Compute new weights as
    		\begin{equation*}
    		w_i^{(r)} \propto \frac{p(\boldsymbol{\theta}_i^{(r)})}{\sum_{i=0}^{m} w_i^{(r-1)}K_r(\boldsymbol{\theta}_i^{(r)} | \boldsymbol{\theta}_i^{(r-1)})} 
    		\end{equation*}
    		\STATE Normalize $w_i^{(r)}$ subject to $\sum_{i = 0}^m w_i^{(r)}$ 
    	\ENDFOR

    \ENDFOR

    \RETURN final particles, $\boldsymbol{\theta}^{(R-1)}$
\end{algorithmic}
\caption{ABCSeq Algorithm} 
\label{ABCSMCAlg} 

\end{algorithm}

\end{minipage}

In Algorithm \ref{ABCSMCAlg}, $K_r(\cdot | \cdot)$ is a conditional density that serves as a transition kernel to move sampled parameters and then appropriately weight the accepted values, which are the parameter valuations which produce trajectories sufficiently close to the data.
In the context of real-valued parameters, which we consider here, $K_r(\boldsymbol{\theta}^* | \boldsymbol{\theta})$ is taken to be a multivariate normal distribution centred near $\boldsymbol{\theta}$.
There are many adaptive schemes to increase the accuracy and the speed of ABCSeq\cite{Del2012,Bonassi2015}, 
which vary from the choice of kernel\cite{Filippi2013}, $K_r(\cdot | \cdot)$ to adapting the discrepancy threshold\cite{Prangle2017}.
We implement the proposed kernel densities presented in \cite{Bonassi2015} and chose an adaptive discrepancy threshold such that $\boldsymbol{\epsilon}_{r+1} = median(\boldsymbol{{\rho}}_r)$, where $\boldsymbol{\rho}_r$ is the vector of all accepted distances for each particle, calculated in line 9 of Algorithm \ref{ABCSMCAlg}.
However, a larger number of particles, $m$, is required than the desired number of independent samples from the ABC posterior with discrepancy threshold $\boldsymbol{\epsilon}$.
For our implementation, we set a maximum number of iterations in the loop in line 3 of Algorithm \ref{ABCSMCAlg} to avoid infinite loops, and we return the particles of the previous sampled parameters if this were to be the case.

\subsection{Probability Computation}

In the final phase of our approach, a probability estimate is computed corresponding to the satisfaction of a CSL specification formula $\phi$ by a system of interest such that $\textbf{S}\models \phi$.
To calculate the probability that the system satisfies the specified property, we require two inputs - the posterior distribution over the whole set of kinetic parameters, $\boldsymbol{\theta}$, discussed in Section \ref{BayesianInf}, and the feasible set of parameters that have been calculated in Sec. \ref{ParamSynthesis}:

\begin{definition}
Given a CSL specification $\phi$ and observed data $D$  from the system $\emph{\textbf{S}}$, the probability that $\emph{\textbf{S}}\models \phi$ is given by
\begin{equation}
\mathbb{C} = P(\emph{\textbf{S}} \models \phi \ | \ D) = \int_{\Theta_{\phi }}p(\boldsymbol{\theta} \ | \ D ) \text{d}\boldsymbol{\theta},
\end{equation}
\end{definition}
where $\Theta_{\phi}$ denotes the feasible set of parameters.
We estimate this integral with the use of Markov chain Monte Carlo (MCMC) methods focusing on the slice sampling technique \cite{Neal2003a}.

\section{Results}
\subsubsection*{Experimental Setup}
All experiments have been run on an Intel(R) Xeon(R) CPU E5-1660 v3 @ 3.00GHz, 16 cores with 16GB memory.
We work with partially observed data of the type discussed in Section \ref{PartiallyObservedData}.
Data is of the form $\textbf{Y}(\tilde{t}_i) = \textbf{X}(\tilde{t}_i) + \boldsymbol{\xi}$, where in the case of noisy observations, the additive noise for each observation $j$  will be given by, $\xi_j \sim \mathcal{N}(0,\sigma)$ and $\sigma = 2$.
The data generating system, $\textbf{S}$ will in fact be a pCTMC with a chosen combination of parameters, of which we consider two.
The first combination, $\boldsymbol{\theta}_{\phi} \in \Theta_{\phi}$, have been chosen such that $\mathcal{M}_{\boldsymbol{\theta}_{\phi}} \models \phi$, that is, the pCTMC model $\mathcal{M}_{\boldsymbol{\theta}_{\phi}}$, governed by $\boldsymbol{\theta}_{\phi}$, satisfies the property of interest.
The second combination we choose are the parameters given by $\boldsymbol{\theta}_{\neg\phi} \in \Theta \setminus \Theta_{\phi}$, such that $\mathcal{M}_{\boldsymbol{\theta}_{\neg\phi}}  \not\models \phi$. 
We will consider the scenario where we have both noisy and noiseless observations.
To summarise, we have instances where we observe either 10 or 20 data points per species, which can be either noisy or noiseless and working with data that has been produced by either $\mathcal{M}_{\boldsymbol{\theta}_{\phi}}$ or $\mathcal{M}_{\boldsymbol{\theta}_{\neg\phi}}$.
To ensure the inference does not depend on the initialisation of the ABCSeq technique, we ran 10 independent batches with 1000 particles each and calculated the corresponding weighted means and variance of the batches to derive the inferred mean and credibility intervals.
The ABCSeq method produces sampled particles from the posterior probability distribution, which we use to calculate the mean, $\boldsymbol{\mu}$ and the covariance, $\boldsymbol{\Sigma}$, of the kinetic parameters.
We assume the parameters are independent of each other, thus the nondiagonal elements of the covariance matrix are equal to 0.
The inferred parameters $\boldsymbol{\tilde{\theta}}$ is thus described by a multivariate normal distribution $\boldsymbol{\tilde{\theta}} \sim \mathcal{N} ( \boldsymbol{\mu}, \boldsymbol{\Sigma})$.

The Bayesian statistical model checking method \cite{Zuliani2009} approach collects sample trajectories from the system, and then determines whether the trajectories satisfy a given property and applies statistical techniques, such as calculation of credibility intervals and hypothesis testing, to decide whether the system satisfies the property or not with a degree of probability.

\subsection{Case Study: Finite-state SIR Model}

We take into account the stochastic epidemic model \cite{Kermack1927}, known alternatively as the SIR model.
Epidemiological models of this type behave largely in the same way as CRNs\cite{Ceska2016}.
The model describes the epidemic dynamics of three types, the susceptible group $(S)$, the infected group $(I)$, and recovered group of individuals $(R)$.
The epidemic dynamics can be described with mass action kinetics:
\begin{equation}
S + I \xrightarrow{k_i} I + I, \ I  \xrightarrow{k_r} R.
\end{equation}
Whenever a susceptible individual $S$ encounters an infected individual $I$, the susceptible individual becomes infected with the rate $k_i$ and infected individuals recover at rate $k_r$.
Letting $S$, $I$ and $R$ represent chemical species instead of groups of individuals, this epidemiological model is the same as a CRN.
From now on we treat the SIR model as a CRN.
This CRN is governed by the parameters $\boldsymbol{\theta} = (k_i, k_r)$, where each state of the CTMC describes the combination of the number of molecules for each species.
The problem we consider is as follows.
We assume  that initially there are 95 molecules of species $S$, 5 molecules of species $I$ and 0 molecules of species $R$, thus, the initial state is  $s_0 = (S_0, I_0,R_0) = (95,5,0)$.
We wish to verify the following property, $\phi = P_{> 0.1}[(I>0)U^{[100,150]}(I=0)]$, i.e. whether, with a probability greater than 0.1, the chemical species $I$ dies out strictly within the interval of $t=100$ and $t=150$ seconds.
The data is produced by both $\mathcal{M}_{\boldsymbol{\theta}_{\phi}}$ and $\mathcal{M}_{\boldsymbol{\theta}_{\neg\phi}}$, where $\boldsymbol{\theta}_{\phi} = (0.002,0.05)$ and $\boldsymbol{\theta}_{\neg\phi} = (0.002,0.18)$.

\begin{figure}
\centering
 \includegraphics[width=90mm]{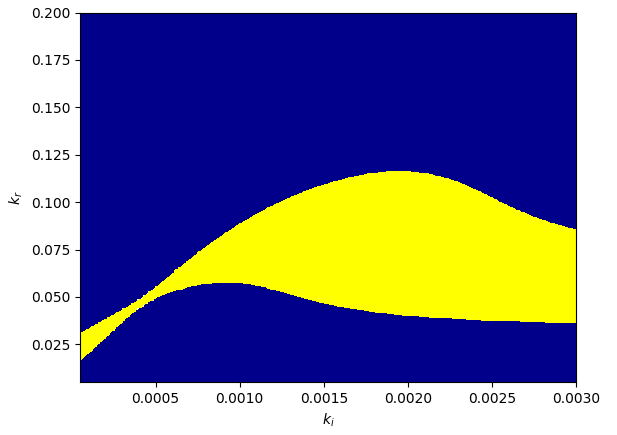}
\caption{ Synthesised Parameter regions are shown here.
The feasible set of parameters, $\mathcal{T}$,  is shown in yellow (lighter colour), meanwhile the infeasible set of parameters, $\mathcal{F}$, is shown in blue (darker colour) $\Theta_{\neg\phi}$, with the undecided areas (if any) shown in white, the set $\mathcal{U}$. }
\label{SIR_ParamSynth}
\end{figure}

In the first phase of our method, we synthesise the feasible set of parameters, $\Theta_{\phi}$.
For the parameter synthesis technique used in our work\cite{Ceska2014,Ceska2016}, we define parameter bounds and we confine our parameters to the set $\Theta = [k_i^{\bot}, k_i^{\top}] \times [k_r^{\bot},k_r^{\top}] = [5\times 10^{-5},0.003] \times [0.005,0.2]$.
The results of the parameter synthesis is shown in Figure \ref{SIR_ParamSynth} and took a total of 3096 seconds (51.6 minutes) to compute.
The second phase of our approach involves learning the kinetic parameters from data, $D$, via the ABCSeq method introduced in Section \ref{BayesianInf}.
To showcase the accuracy of our method, we consider different data scenarios.
We take into account observed data where the observations are either distorted or not by additive noise, and the aforementioned two cases but with additional observed data points.
A full list of different data scenarios and corresponding inferred parameters can be seen in Table \ref{tab:InferredParams}.
As expected, if we observe more, noiseless data points then our inferred parameters converge to the true parameters, $\boldsymbol{\theta}_{\phi}$ or $\boldsymbol{\theta}_{\neg \phi}$.
The accuracy decreases drastically for data produced by the model $\mathcal{M}_{\boldsymbol{\theta}_{\neg \phi}}$.
This is due to the largely uninformative observations as the samples reach steady state.
To increase accuracy, more observations should be taken during the transient period of the model.

\begin{figure}
\vspace{-0.5cm}
\centering
\begin{minipage}{\textwidth}
\includegraphics[width = 120mm]{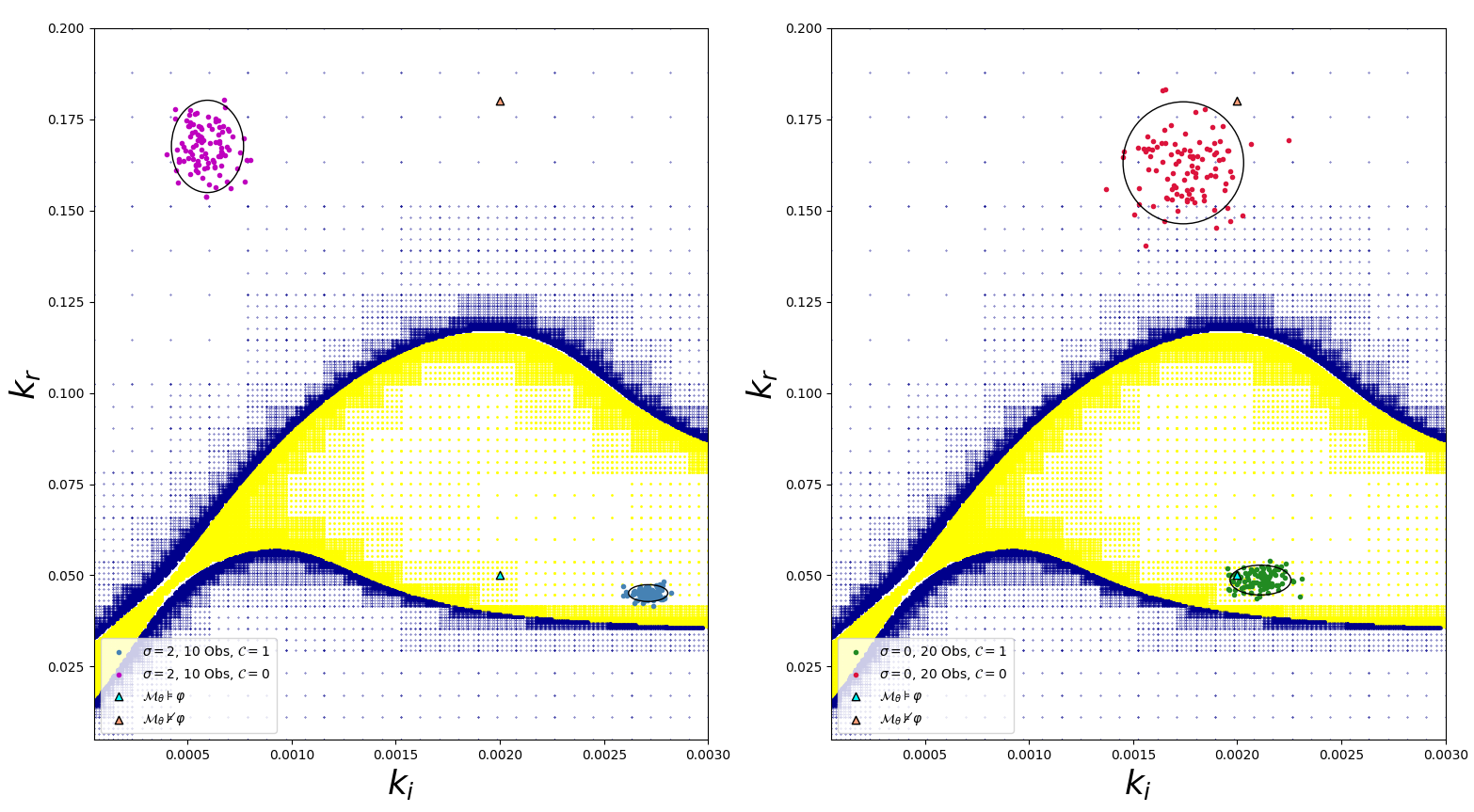}
\caption{Bayesian statistical model checking is performed over the inferred parameters for the case with 10 noisy observations on the left figure and for the case with 20 noiseless observations on the right figure.
Dark blue (left figure) and green points (right figure) represent the probability values of 1 meanwhile the purple (left figure) and red points (right figure) represent probability values of 0.
The parameters chosen to produce the data are represented by the cyan and orange triangles.
The black elliptical lines represent the 95\% credibility intervals of the inferred posterior distribution.
The yellow (or blue) points represent the parameter valuation regions that satisfy ($\mathcal{T}$) (or don't satisfy, $\mathcal{F}$) the property.
}
\label{BayesianSMC}
\end{minipage}
\end{figure}

Bayesian SMC requires multiple simulated trajectories over a given model $\mathcal{M}_{\boldsymbol{\theta}}$ to determine whether $\mathcal{M}_{\boldsymbol{\theta}} \models \phi$.
The issue with Bayesian SMC is that it considers a single instance of parameters, $\boldsymbol{\theta}^0$ and produces multiple simulations and statistically verifies whether the property is satisfied or not.
When inferring parameters, we compute a probability distribution over the set of inferred parameters.
If this distribution were to have a high variance, one would need to sample many parameters from the posterior distribution to sufficiently cover the space of the parameter probability distribution and then produce simulations for Bayesian SMC to evaluate each instantiation of the parameters.
Meanwhile in our approach, we would only need to integrate the posterior distribution $p(\boldsymbol{\theta} | D)$  over the feasible parameter set $\Theta_{\phi}$ to obtain a probability whether this property is satisfied or not.
Bayesian SMC is illustrated in Figure \ref{BayesianSMC}.
For both Bayesian SMC and our method, we first had to infer the parameters to obtain a posterior probability distribution, which in this case, is a bivariate normal distribution.
For Bayesian SMC, we sampled 100 independent evaluations of the parameters, and produced 1000 simulations for each evaluation to determine the probability that the model, $\mathcal{M}_{\boldsymbol{\theta}}$, satisfies the property of interest.
The sampled parameters are represented by the points represented by circles in Figure \ref{BayesianSMC}, meanwhile the 95\% credibility interval for the inferred parameters are represented by the black ellipses.
The computation time for the Bayesian SMC approach was 756 seconds (12.6 minutes). 
With our approach, we simply need to integrate the bivariate normal distribution over the feasible parameter regions over the parameter regions to obtain the values in Table \ref{tab:InferredParams}, and we do this numerically via slice sampling \cite{Neal2003a}.
Both our technique and Bayesian SMC are in agreement, but for the Bayesian SMC approach we would require a larger number of sampled parameters to verify whether or not the entire posterior probability distribution lies in these feasible regions.
Despite the fact that the parameter synthesis for the whole region takes longer to compute, the exhaustive parameter synthesis technique provides us a picture of the whole parameter space which is useful for further experiments and can be done entirely offline.
For our multivariate slice sampler, we heuristically chose the number of samples to be 10000 and the scale estimates for each parameter, $\nu_i$ was chosen to be $\nu_i = 2$, with the initial value of the slice equal to the mean of the inferred posterior probability  distribution.
For  further details on multivariate slice sampling (which leads to the credibility calculation given in Table \ref{tab:InferredParams}), see \cite{Neal2003a}.
For convergence results with statistical guarantees, we refer the reader to \cite{Cowles1996} meanwhile if interested in obtaining an upper bound on the probability calculated, we refer to \cite{Hoeffding1962}.
Both the inference and Bayesian SMC techniques break down if simulating traces for CRNs is costly.
Fortunately, there is ongoing research on approximation techniques that sacrifice the accuracy of Gillespie's algorithm for speed (such as the classical tau-leaping method \cite{Gillespie2001}).
For alternative approximation techniques, see \cite{Higham2008,Warne2019} for more details.
\vspace{-0.5cm}
\section{Conclusions and Further Work}
We have presented a data-driven approach for the verification of CRNs modelled as pCTMCs.
The framework proposed integrates Bayesian inference and formal verification and proves to be a viable alternative to Bayesian SMC methods.
We demonstrate how to infer parameters using noisy and discretely observed data using ABC and with the inferred posterior probability distribution of the parameters at hand, we calculate the probability that the underlying data generating system satisfies a property by integrating over the synthesised feasible parameter regions.
Thus, given data from an underlying system, we can quantitatively assert whether properties of the underlying system are satisfied or not.
Our method differs from that of typical Bayesian SMC as we calculate a single probability value with respect to the entire posterior distribution, meanwhile with Bayesian SMC, we would have to sample a sufficient amount of parameter values to cover the posterior distribution, thereon generate traces to determine whether a property is satisfied or not.
Future work consists of integrating both learning and verification further as is done in \cite{Bortolussi2018} to improve the scalability of the parameter synthesis, working with different model classes such as stochastic differential equations \cite{Gardiner1985,Gillespie2000} and models with actions, as is done in \cite{Polgreen2017a}.

\begin{table}
\centering
\resizebox{0.83\textwidth}{!}
{
\begin{minipage}{\textwidth}
\begin{tabular}{ |p{1.5cm}||p{1.0cm}||p{2cm}||p{2.2cm}||p{1cm}||p{1.5cm}|  }
 \hline
 \multicolumn{6}{|c|}{\textbf{Inferred Parameters}} \\
 \hline
 \textbf{Data} & \textbf{ True par.}  & \textbf{Mean} & \textbf{Std. Dev.}
 &\textbf{Prob.}& \textbf{Comp. Time (s)}\\
 \hline
 10 Obs. with Noise   & \ \ \    $\boldsymbol{\theta}_{\phi}$ &$\mu_{k_i} = 0.0027$& $\sigma_{k_i} = 4.7\times 10^{-5} $ & \ \ \ 1 & 11791 \\ 
 & & $\mu_{k_r} = 0.0451$ & $\sigma_{k_r} = 0.0012 $& &   \\
\cline{2-6}
 				 	  & \ \ $\boldsymbol{\theta}_{\neg\phi}$ & $\mu_{k_i} = 0.0006$& $\sigma_{k_i} = 8.7\times 10^{-5}
 $ & \ \ \  0 & 75  \\
 					  &   & $\mu_{k_r} = 0.1676 $   &$\sigma_{k_r} = 0.0063
 $&  &   \\
 \hline
 20 Obs. with Noise & \ \ \ $\boldsymbol{\theta}_{\phi}$ & $\mu_{k_i} = 0.0022$& $\sigma_{k_i} = 0.0001 $ & \ \ \ \ 0.9901
 & 10840\\
 				 	 &      &  $\mu_{k_r} =  0.0468$&$\sigma_{k_r} = 0.0036$ &  &  \\
 \cline{2-6}
 				 	  & \ \  $\boldsymbol{\theta}_{\neg\phi}$ & $\mu_{k_i} = 0.0015$& $\sigma_{k_i} =0.0002$ & \ \ \  0 & 3256 \\
 					  &   & $\mu_{k_r} =0.1620 $&$\sigma_{k_r} = 0.0104
 $&  &   \\
 \hline
 10 Obs without Noise & \ \  $\boldsymbol{\theta}_{\phi}$ & $\mu_{k_i} = 0.0019$& $\sigma_{k_i} =8.8\times 10^{-5}
 $ & \ \ \ \  0.9969 & 7585\\
  &      &  $\mu_{k_r} = 0.0549$& $\sigma_{k_r} = 0.00514 $ & &  \\
 \cline{2-6}
 				 	  & \ \ $\boldsymbol{\theta}_{\neg\phi}$ & $\mu_{k_i} = 0.0015$& $\sigma_{k_i} = 0.0001 $ & \ \ \  0 & 3802 \\
 					  &   & $\mu_{k_r} = 0.1565 $   &$\sigma_{k_r} = 0.0074
 $& &  \\
 \hline
 20 Obs. without Noise & \ \ \  $ \boldsymbol{\theta}_{\phi}$ & $\mu_{k_i} = 0.0021$& $\sigma_{k_i} = 7.3\times 10^{-5}
 $ & \ \ \  1 & 15587 \\
 				 	 &      &  $\mu_{k_r} = 0.0487$ & $\sigma_{k_r} = 0.0020 $ &  &  \\
 \cline{2-6}
 				 	  & \ \  $\boldsymbol{\theta}_{\neg\phi}$ & $\mu_{k_i} = 0.0017$&$\sigma_{k_i} = 0.0001 $ & \ \ \  0 & 5194  \\
 					  &   & $\mu_{k_r} = 0.1630$&$\sigma_{k_r} = 0.0084 $& &      \\
 \hline 
\end{tabular}

\caption{We have four different data scenarios to produce the data.
Data within each of these datasets is produced given a combination of parameters that satisfy the property of interest, $\boldsymbol{\theta}_{\phi} = (0.002,0.05)$ and those that do not satisfy the property of interest $\boldsymbol{\theta}_{\neg \phi} = (0.002,0.18)$,  (with or without additive noise $\sigma = 2$).
We integrate the corresponding posterior distribution to give us the probability in column 5.  
\label{tab:InferredParams}
}
\end{minipage}
}
\end{table}
\newpage
\bibliographystyle{splncs04}
 \bibliography{GWM}

\begin{thebibliography}{10}
\providecommand{\url}[1]{\texttt{#1}}
\providecommand{\urlprefix}{URL }
\providecommand{\doi}[1]{https://doi.org/#1}

\bibitem{Agha2018}
Agha, G., Palmskog, K.: A survey of statistical model checking. ACM Trans.
  Model. Comput. Simul.  \text{28}(1),  6:1--6:39 (Jan 2018)

\bibitem{Andrieu2009}
Andrieu, C., Roberts, G.O., et~al.: The pseudo-marginal approach for efficient
  monte carlo computations. The Annals of Statistics  \text{37}(2),  697--725
  (2009)

\bibitem{Angeli2009}
Angeli, D.: A tutorial on chemical reaction network dynamics. European Journal
  of Control  \text{15}(3),  398 -- 406 (2009)

\bibitem{Aziz1996}
Aziz, A., Sanwal, K., Singhal, V., Brayton, R.: Verifying continuous time
  {M}arkov chains. In: Alur, R., Henzinger, T.A. (eds.) Computer Aided
  Verification. pp. 269--276. Springer Berlin Heidelberg, Berlin, Heidelberg
  (1996)

\bibitem{Baier2003}
Baier, C., Haverkort, B.R., Hermanns, H., Katoen, J.: Model-checking algorithms
  for continuous-time {Markov} chains. {IEEE} Trans. Software Eng.
  \text{29}(6),  524--541 (2003)

\bibitem{Baier:2008:PMC:1373322}
Baier, C., Katoen, J.: Principles of model checking. {MIT} Press (2008)

\bibitem{Barnes2011a}
Barnes, C.P., Silk, D., Sheng, X., Stumpf, M.P.: Bayesian design of synthetic
  biological systems. Proceedings of the National Academy of Sciences
  \text{108}(37),  15190--15195 (2011)

\bibitem{Barnes2011b}
Barnes, C.P., Silk, D., Stumpf, M.P.: Bayesian design strategies for synthetic
  biology. Interface focus  \text{1}(6),  895--908 (2011)

\bibitem{Beaumont2010}
Beaumont, M.A.: Approximate bayesian computation in evolution and ecology.
  Annual review of ecology, evolution, and systematics  \text{41},  379--406
  (2010)

\bibitem{Beaumont2009}
Beaumont, M.A., Cornuet, J.M., Marin, J.M., Robert, C.P.: Adaptive approximate
  bayesian computation. Biometrika  \text{96}(4),  983--990 (2009)

\bibitem{Beaumont2002}
Beaumont, M.A., Zhang, W., Balding, D.J.: Approximate bayesian computation in
  population genetics. Genetics  \text{162}(4),  2025--2035 (2002)

\bibitem{Bonassi2015}
Bonassi, F.V., West, M., et~al.: Sequential monte carlo with adaptive weights
  for approximate bayesian computation. Bayesian Analysis  \text{10}(1),
  171--187 (2015)

\bibitem{Bortolussi2016}
Bortolussi, L., Milios, D., Sanguinetti, G.: Smoothed model checking for
  uncertain continuous-time {M}arkov chains. Inf. Comput.  \text{247}(C),
  235--253 (Apr 2016)

\bibitem{Bortolussi2013}
Bortolussi, L., Sanguinetti, G.: Learning and designing stochastic processes
  from logical constraints. In: International Conference on Quantitative
  Evaluation of Systems. pp. 89--105. Springer (2013)

\bibitem{Bortolussi2018}
Bortolussi, L., Silvetti, S.: {B}ayesian statistical parameter synthesis for
  linear temporal properties of stochastic models. In: Beyer, D., Huisman, M.
  (eds.) Tools and Algorithms for the Construction and Analysis of Systems. pp.
  396--413. Springer International Publishing, Cham (2018)

\bibitem{Box1973}
Box, G., Tiao, G.: {B}ayesian Inference in Statistical Analysis. Wiley Classics
  Library, Wiley (1973)

\bibitem{Boys2008}
Boys, R.J., Wilkinson, D.J., Kirkwood, T.B.: Bayesian inference for a
  discretely observed stochastic kinetic model. Statistics and Computing
  \text{18}(2),  125--135 (2008)

\bibitem{Brim2013}
Brim, L., {\v{C}}e{\v{s}}ka, M., Dra{\v{z}}an, S., {\v{S}}afr{\'a}nek, D.:
  Exploring parameter space of stochastic biochemical systems using
  quantitative model checking. In: Sharygina, N., Veith, H. (eds.) Computer
  Aided Verification. pp. 107--123. Springer Berlin Heidelberg, Berlin,
  Heidelberg (2013)

\bibitem{Broemeling2017}
Broemeling, L.: Bayesian Inference for Stochastic Processes. CRC Press (2017)

\bibitem{Ceska2014}
Ceska, M., Dannenberg, F., Paoletti, N., Kwiatkowska, M., Brim, L.: Precise
  parameter synthesis for stochastic biochemical systems. Acta Inf.
  \text{54}(6),  589--623 (2014)

\bibitem{Ceska2016}
Ceska, M., Pilar, P., Paoletti, N., Brim, L., Kwiatkowska, M.Z.: {PRISM-PSY:}
  precise gpu-accelerated parameter synthesis for stochastic systems. In: Tools
  and Algorithms for the Construction and Analysis of Systems - 22nd
  International Conference, {TACAS} 2016, Held as Part of the European Joint
  Conferences on Theory and Practice of Software, {ETAPS} 2016, Eindhoven, The
  Netherlands, April 2-8, 2016, Proceedings. pp. 367--384 (2016)

\bibitem{Cook2009}
Cook, M., Soloveichik, D., Winfree, E., Bruck, J.: Programmability of chemical
  reaction networks pp. 543--584 (2009)

\bibitem{Cowles1996}
Cowles, M.K., Carlin, B.P.: Markov chain monte carlo convergence diagnostics: a
  comparative review. Journal of the American Statistical Association
  \text{91}(434),  883--904 (1996)

\bibitem{Del2012}
Del~Moral, P., Doucet, A., Jasra, A.: An adaptive sequential monte carlo method
  for approximate bayesian computation. Statistics and Computing  \text{22}(5),
   1009--1020 (2012)

\bibitem{Filippi2013}
Filippi, S., Barnes, C.P., Cornebise, J., Stumpf, M.P.: On optimality of
  kernels for approximate bayesian computation using sequential monte carlo.
  Statistical applications in genetics and molecular biology  \text{12}(1),
  87--107

\bibitem{Galagali2014}
Galagali, N., Marzouk, Y.M.: Bayesian inference of chemical kinetic models from
  proposed reactions. Chemical Engineering Science  (2015)

\bibitem{Gardiner1985}
Gardiner, C.: Stochastic Methods: A Handbook for the Natural and Social
  Sciences. 13, Springer-Verlag Berlin Heidelberg, 4 edn.

\bibitem{Georgoulas2017}
Georgoulas, A., Hillston, J., Sanguinetti, G.: Unbiased {Bayesian} inference
  for population {Markov} jump processes via random truncations. Statistics and
  Computing  \text{27}(4),  991--1002 (2017)

\bibitem{Gillespie1977b}
Gillespie, D.T.: Exact stochastic simulation of coupled chemical reactions. The
  Journal of Physical Chemistry  \text{81}(25),  2340--2361 (1977)

\bibitem{Gillespie1992}
Gillespie, D.T.: A rigorous derivation of the chemical master equation. Physica
  A: Statistical Mechanics and its Applications  \text{188}(1),  404 -- 425
  (1992)

\bibitem{Gillespie2001}
Gillespie, D.T.: Approximate accelerated stochastic simulation of chemically
  reacting systems. The Journal of Chemical Physics  \text{115}(4),  1716--1733
  (2001)

\bibitem{Gillespie2000}
Gillespie, D.T., Gillespie, D.T.: {The chemical Langevin equation The chemical
  Langevin equation}  \text{297}(2000) (2000)

\bibitem{Golightly2006}
Golightly, A., Wilkinson, D.J.: Bayesian sequential inference for stochastic
  kinetic biochemical network models. Journal of Computational Biology
  \text{13}(3),  838--851 (2006)

\bibitem{Golighty2011}
Golightly, A., Wilkinson, D.J.: Bayesian parameter inference for stochastic
  biochemical network models using particle markov chain monte carlo. Interface
  focus  \text{1}(6),  807--820 (2011)

\bibitem{Golighty2015}
Golightly, A., Wilkinson, D.J.: Bayesian inference for markov jump processes
  with informative observations. Statistical applications in genetics and
  molecular biology  \text{14}(2),  169--188 (2015)

\bibitem{Gyori2014}
Gyori, B.M., Paulin, D., Palaniappan, S.K.: Probabilistic verification of
  partially observable dynamical systems. arXiv preprint arXiv:1411.0976
  (2014)

\bibitem{Haesaert2015}
Haesaert, S., den Hof, P.M.J.V., Abate, A.: Data-driven and model-based
  verification: a {Bayesian} identification approach. CoRR
  \text{abs/1509.03347} (2015)

\bibitem{Han2008}
Han, T., Katoen, J.P., Mereacre, A.: Approximate parameter synthesis for
  probabilistic time-bounded reachability. 2008 Real-Time Systems Symposium pp.
  173--182 (2008)

\bibitem{Higham2008}
Higham, D.J.: Modeling and simulating chemical reactions. SIAM Review
  \text{50}(2),  347--368 (May 2008)

\bibitem{Hoeffding1962}
Hoeffding, W.: Probability inequalities for sums of bounded random variables
  (1962)

\bibitem{Zuliani2009}
Jha, S.K., Clarke, E.M., Langmead, C.J., Legay, A., Platzer, A., Zuliani, P.: A
  {B}ayesian approach to model checking biological systems. In: Degano, P.,
  Gorrieri, R. (eds.) Computational Methods in Systems Biology. pp. 218--234.
  Springer Berlin Heidelberg, Berlin, Heidelberg (2009)

\bibitem{Karlin1975}
Karlin, S., Taylor, H., Taylor, H., Taylor, H., Collection, K.M.R.: A First
  Course in Stochastic Processes. No.~v. 1, Elsevier Science (1975)

\bibitem{Kermack1927}
Kermack, W.: A contribution to the mathematical theory of epidemics.
  Proceedings of the Royal Society of London A: Mathematical, Physical and
  Engineering Sciences  \text{115}(772),  700--721 (1927)

\bibitem{Kwiatkowska2007}
Kwiatkowska, M., Norman, G., Parker, D.: Stochastic model checking. In:
  Bernardo, M., Hillston, J. (eds.) Formal Methods for the Design of Computer,
  Communication and Software Systems: Performance Evaluation (SFM'07). LNCS
  (Tutorial Volume), vol.~4486, pp. 220--270. Springer (2007)

\bibitem{Kwiatkowska2014}
Kwiatkowska, M., Thachuk, C.: Probabilistic model checking for biology. In:
  Software Safety and Security. NATO Science for Peace and Security Series - D:
  Information and Communication Security, IOS Press (2014)

\bibitem{Kwiatkowska2017}
Kwiatkowska, M., Norman, G., Parker, D.: Probabilistic model checking: advances
  and applications. In: Drechsler, R. (ed.) Formal System Verification:
  State-of the-Art and Future Trends, pp. 73--121. Springer Verlag (2017)

\bibitem{Kwiatkowska2011}
Kwiatkowska, M.Z., Norman, G., Parker, D.: {PRISM} 4.0: Verification of
  probabilistic real-time systems. In: Computer Aided Verification - 23rd
  International Conference, {CAV} 2011, Snowbird, UT, USA, July 14-20, 2011.
  Proceedings. pp. 585--591 (2011)

\bibitem{Kypraios2017}
Kypraios, T., Neal, P., Prangle, D.: A tutorial introduction to bayesian
  inference for stochastic epidemic models using approximate bayesian
  computation. Mathematical Biosciences  \text{287},  42 -- 53 (2017). 50th
  Anniversary Issue

\bibitem{Lawrence2010}
Lawrence, N.D., Girolami, M., Rattray, M., Sanguinetti, G. (eds.): Learning and
  Inference in Computational Systems Biology. MIT Press, Cambridge,
  Massachusetts ; London (2010)

\bibitem{Liepe2013}
Liepe, J., Filippi, S., Komorowski, M., Stumpf, M.P.H.: Maximizing the
  information content of experiments in systems biology. PLOS Computational
  Biology  \text{9}(1),  1--13 (01 2013)

\bibitem{Milios2017}
Milios, D., Sanguinetti, G., Schnoerr, D.: Probabilistic model checking for
  continuous-time markov chains via sequential bayesian inference. In: McIver,
  A., Horvath, A. (eds.) Quantitative Evaluation of Systems. pp. 289--305.
  Springer International Publishing, Cham (2018)

\bibitem{Murphy2012}
Murphy, K.P.: Machine learning - a probabilistic perspective. Adaptive
  computation and machine learning series, {MIT} Press (2012)

\bibitem{Neal2003a}
Neal, R.M.: Slice sampling. Ann. Statist.  \text{31}(3),  705--767 (06 2003)

\bibitem{Polgreen2016a}
Polgreen, E., Wijesuriya, V.B., Haesaert, S., Abate, A.: Data-efficient
  {Bayesian} verification of parametric {Markov} chains. In: Quantitative
  Evaluation of Systems - 13th International Conference, {QEST} 2016, Quebec
  City, QC, Canada, August 23-25, 2016, Proceedings. pp. 35--51 (2016)

\bibitem{Polgreen2017a}
Polgreen, E., Wijesuriya, V.B., Haesaert, S., Abate, A.: Automated experiment
  design for data-efficient verification of parametric {Markov} decision
  processes. In: Quantitative Evaluation of Systems - 14th International
  Conference, {QEST} 2017, Berlin, Germany, September 5-7, 2017, Proceedings.
  pp. 259--274 (2017)

\bibitem{Prangle2017}
Prangle, D., et~al.: Adapting the abc distance function. Bayesian Analysis
  \text{12}(1),  289--309 (2017)

\bibitem{Rasmussen2003}
Rasmussen, C.E.: Gaussian processes in machine learning. In: Summer School on
  Machine Learning. pp. 63--71. Springer (2003)

\bibitem{Revell2018}
Revell, J., Zuliani, P.: Stochastic rate parameter inference using the
  cross-entropy method. In: {\v{C}}e{\v{s}}ka, M., {\v{S}}afr{\'a}nek, D.
  (eds.) Computational Methods in Systems Biology. pp. 146--164. Springer
  International Publishing, Cham (2018)

\bibitem{Sanguinetti2006}
Sanguinetti, G., Lawrence, N.D., Rattray, M.: {Probabilistic inference of
  transcription factor concentrations and gene-specific regulatory activities}.
  Bioinformatics  \text{22}(22),  2775--2781 (09 2006)

\bibitem{Schnoerr2017}
Schnoerr, D., Sanguinetti, G., Grima, R.: Approximation and inference methods
  for stochastic biochemical kinetics: a tutorial review. Journal of Physics A:
  Mathematical and Theoretical  \text{50}(9),  093001 (2017)

\bibitem{Sisson2018}
Sisson, S.A., Fan, Y., Beaumont, M.: Handbook of approximate Bayesian
  computation. Chapman and Hall/CRC (2018)

\bibitem{Sisson2007}
Sisson, S.A., Fan, Y., Tanaka, M.M.: Sequential monte carlo without
  likelihoods. Proceedings of the National Academy of Sciences  \text{104}(6),
  1760--1765 (2007)

\bibitem{Toni2008}
Toni, T., Welch, D., Strelkowa, N., Ipsen, A., Stumpf, M.P.: Approximate
  bayesian computation scheme for parameter inference and model selection in
  dynamical systems. Journal of the Royal Society Interface  \text{6}(31),
  187--202 (2008)

\bibitem{Vanlier2014}
Vanlier, J., Tiemann, C.A., Hilbers, P.A., van Riel, N.A.: Optimal experiment
  design for model selection in biochemical networks. BMC systems biology
  \text{8}(1), ~20 (2014)

\bibitem{Warne2019}
Warne, D.J., Baker, R.E., Simpson, M.J.: Simulation and inference algorithms
  for stochastic biochemical reaction networks: from basic concepts to
  state-of-the-art. Journal of the Royal Society Interface  \text{16}(151),
  20180943 (2019)

\bibitem{Wilkinson2010}
Wilkinson, D.J.: Parameter inference for stochastic kinetic models of bacterial
  gene regulation: a bayesian approach to systems biology. In: Proceedings of
  9th Valencia International Meeting on Bayesian Statistics. pp. 679--705
  (2010)

\bibitem{Wilkinson2011}
Wilkinson, D.: Stochastic Modelling for Systems Biology, Second Edition.
  Chapman \& Hall/CRC Mathematical and Computational Biology, Taylor \& Francis
  (2011)

\bibitem{Wilkinson2013}
Wilkinson, R.D.: Approximate bayesian computation (abc) gives exact results
  under the assumption of model error. Statistical applications in genetics and
  molecular biology  \text{12}(2),  129--141 (2013)

\bibitem{Woods2016}
Woods, M.L., Leon, M., Perez-Carrasco, R., Barnes, C.P.: A statistical approach
  reveals designs for the most robust stochastic gene oscillators. ACS
  synthetic biology  \text{5}(6),  459--470 (2016)

\bibitem{Zuliani2015}
Zuliani, P.: Statistical model checking for biological applications.
  International Journal on Software Tools for Technology Transfer
  \text{17}(4),  527--536 (Aug 2015)

\bibitem{Zuliani2010}
Zuliani, P., Platzer, A., Clarke, E.M.: {Bayesian} statistical model checking
  with application to {Stateflow}/{Simulink} verification. vol.~43, pp.
  338--367 (2013)

\end{thebibliography}
 
\end{document}